\documentclass[10pt, conference, final]{IEEEtran}
\IEEEoverridecommandlockouts
\usepackage{amsmath,amssymb,amsfonts} \usepackage{graphicx} \usepackage{listings} \usepackage{enumitem} \usepackage{ulem} \usepackage{soul} \usepackage{algpseudocode}  \usepackage{algorithm}  
\usepackage{tcolorbox} \usepackage{tabularx}  \usepackage{booktabs}  \usepackage{multirow}  \usepackage{url}  \usepackage{ulem} 
\algrenewcommand\algorithmicrequire{\textbf{Input:}}
\algrenewcommand\algorithmicensure{\textbf{Output:}}

\newcommand{\PreserveBackslash}[1]{\let\temp=\\#1\let\\=\temp}
\newcolumntype{C}[1]{>{\PreserveBackslash\centering}p{#1}}
\newcolumntype{R}[1]{>{\PreserveBackslash\raggedleft}p{#1}}
\newcolumntype{L}[1]{>{\PreserveBackslash\raggedright}p{#1}}

\newcommand*{\algrule}[1][\algorithmicindent]{\hspace*{.5em}\vrule\vrule
width 0pt height \baselineskip depth .25\baselineskip\hspace*{\dimexpr#1-.5em}}
\makeatletter
\newcount\ALG@printindent@tempcnta
\def\ALG@printindent{    \ifnum \theALG@nested>0    \ifx\ALG@text\ALG@x@notext        \else
    \unskip
        \ALG@printindent@tempcnta=1
    \loop
    \algrule[\csname ALG@ind@\the\ALG@printindent@tempcnta\endcsname]    \advance \ALG@printindent@tempcnta 1
    \ifnum \ALG@printindent@tempcnta<\numexpr\theALG@nested+1\relax    \repeat
    \fi
    \fi
}\usepackage{etoolbox}
\patchcmd{\ALG@doentity}{\noindent\hskip\ALG@tlm}{\ALG@printindent}{}{\errmessage{failed to patch}}
\makeatother
\algtext*{EndWhile}\algtext*{EndFor}\algtext*{EndIf}

\algblockdefx[ForB]{ForB}{EndForB}[0]{\textbf{for} }{$)$}
\algcblockdefx[DoB]{ForB}{DoB}{EndDoB}{\textbf{do}}{\textbf{end for}}
\algtext*{EndDoB}
\definecolor{forestgreen}{HTML}{017501}
\definecolor{orangep}{rgb}{0.71, 0.43, 0.89}
\definecolor{orp}{rgb}{1, 0.5, 0.2}
\definecolor{dkgreen}{rgb}{0,0.6,0}
\definecolor{gray}{rgb}{0.3,0.3,0.3}
\definecolor{mauve}{rgb}{0.58,0,0.82}
\definecolor{PastelDark}{HTML}{FF9999}
\definecolor{PastelDark2}{HTML}{ff99e6}
\definecolor{PastelBlue}{HTML}{BBCCEE}
\definecolor{PastelGreen}{HTML}{CCDDAA}
\definecolor{PastelOrange}{HTML}{EEEEBB}
\definecolor{PastelRed}{HTML}{FFCCCC}
\definecolor{PastelPurple}{HTML}{CCEEFF}
\definecolor{PastelCyan}{HTML}{CCEEFF}
\definecolor{GreyHL}{rgb}{0.9,0.9,0.9}

\definecolor{pinka1}{HTML}{DF8B86}
\definecolor{pinki2}{HTML}{F8CECC}
\definecolor{greenc1}{HTML}{61CC61}
\definecolor{blueo1}{HTML}{9ABBFE}

\definecolor{hg}{HTML}{CCDDAA} \definecolor{hb}{HTML}{CCEEFF} \definecolor{hr}{HTML}{FFCCCC} \definecolor{hy}{HTML}{FFF2CC} \definecolor{hp}{HTML}{E1D5E7}

\newtcolorbox{resultbox}{
    colback=black!5!white,
    colframe=white!50!black, 
    boxsep=0mm,
}

\newcommand{\tool}{\textsc{SynthFuzz}{}}

\DeclareMathOperator{\synth}{SYNTH}
\DeclareMathOperator{\synthlocate}{LOCATE}
\DeclareMathOperator{\synthmatch}{MATCH}
\DeclareMathOperator{\instantiate}{INSTANTIATE}

\DeclareMathOperator{\getLeftSibling}{getLeft}
\DeclareMathOperator{\getRightSibling}{getRight}

\DeclareMathOperator{\getParent}{getParent}
\DeclareMathOperator{\pName}{pName}

\DeclareMathOperator{\getNext}{getNext}

\newcommand{\avgDivImprove}{1.75$\times$}
\newcommand{\avgDivImproveGrammar}{1.75$\times$}
\newcommand{\avgDivImproveMLIRSmith}{4.60$\times$}
\newcommand{\avgDivImproveNeuRI}{5.56$\times$}

\newcommand{\avgConDivImprove}{1.70$\times$}

\newcommand{\avgConDivImproveGrammar}{1.70$\times$}
\newcommand{\avgConDivImproveMLIRSmith}{4.16$\times$}
\newcommand{\avgConDivImproveNeuRI}{5.32$\times$}

\newcommand{\avgDataDivImprove}{1.43$\times$}
\newcommand{\avgDataDivImproveGrammar}{1.88$\times$}
\newcommand{\avgDataDivImproveMLIRSmith}{4.38$\times$}
\newcommand{\avgDataDivImproveNeuRI}{4.18$\times$}

\newcommand{\newConPairs}{99}
\newcommand{\newDataPairs}{57}

\newcommand{\avgCovImprove}{1.22$\times$}
\newcommand{\avgCovImproveGrammar}{1.22$\times$}
\newcommand{\avgCovImproveMLIRSmith}{29.78$\times$}
\newcommand{\avgCovImproveNeuRI}{17.47$\times$}

\newcommand{\avgCovImproveGrammarMLIR}{1.51$\times$}
\newcommand{\avgCovImproveMLIRSmithMLIR}{1.99$\times$}
\newcommand{\avgCovImproveNeuRIMLIR}{9.06$\times$}

\newcommand{\avgCovImproveGrammarCIRCT}{1.21$\times$}
\newcommand{\avgCovImproveMLIRSmithCIRCT}{43.25$\times$}
\newcommand{\avgCovImproveNeuRICIRCT}{34.50$\times$}

\newcommand{\maxContextValidImprove}{1.11$\times$}
\newcommand{\kFourValidImprove}{1.11$\times$}
\newcommand{\lFourValidImprove}{1.07$\times$}
\newcommand{\rFourValidImprove}{1.03$\times$}

\newcommand{\paramGenMLIRReduction}{0.57$\times$}

\newcommand{\avgMLIRSmithLOCPerDialect}{447}
\newcommand{\MLIRSmithTotalLOC}{11,434}

\newcommand{\hlcolor}[2]{{\sethlcolor{#1}\hl{#2}}}

\lstdefinelanguage{mlir}{
    sensitive=false, 
  morecomment=[l]{//}, 
}
\makeatletter
\newcommand{\newlineauthors}{  \end{@IEEEauthorhalign}\hfill\mbox{}\par
  \mbox{}\hfill\begin{@IEEEauthorhalign}
}
\makeatother

\author{\IEEEauthorblockN{Ben Limpanukorn}
\IEEEauthorblockA{University of California, Los Angeles \\
blimpan@cs.ucla.edu}
\and
\IEEEauthorblockN{Jiyuan Wang}
\IEEEauthorblockA{University of California, Los Angeles \\
wangjiyuan@cs.ucla.edu}
\and
\IEEEauthorblockN{Hong Jin Kang}
\IEEEauthorblockA{University of California, Los Angeles \\
hjkang@cs.ucla.edu}
\newlineauthors
\IEEEauthorblockN{Zitong Zhou}
\IEEEauthorblockA{University of California, Los Angeles \\
zitongzhou@cs.ucla.edu}
\and
\IEEEauthorblockN{Miryung Kim}
\IEEEauthorblockA{University of California, Los Angeles \\
miryung@cs.ucla.edu}
}

\begin{document}

\title{Fuzzing MLIR Compilers with Custom Mutation Synthesis\\
}

\maketitle

\begin{abstract}

Compiler technologies in deep learning and domain-specific hardware acceleration are increasingly adopting extensible compiler frameworks such as Multi-Level Intermediate Representation (MLIR) to facilitate more efficient development.
With MLIR, compiler developers can easily define their own custom IRs in the form of MLIR dialects.
However, the diversity and rapid evolution of such custom IRs make it impractical to manually write a custom test generator for each dialect.

To address this problem, we design a new test generator called \tool{} that combines grammar-based fuzzing with custom mutation synthesis. The key essence of \tool{} is two fold:
(1) It automatically infers parameterized context-dependent custom mutations from existing test cases.
(2) It then concretizes the mutation's content depending on the target context and
reduces the chance of inserting invalid edits by performing $k$-ancestor and prefix/postfix matching. It obviates the need to manually define custom mutation operators for each dialect.

We compare \tool{} to three baselines: Grammarinator---a grammar-based fuzzer without custom mutations, MLIRSmith---a custom test generator for MLIR core dialects,
and NeuRI---a custom test generator for ML models with parameterization of tensor shapes.
We conduct this comprehensive comparison on four different MLIR projects. 
Each project defines a new set of MLIR dialects where manually writing a custom test generator would take weeks of effort.
Our evaluation shows that \tool{} on average improves MLIR dialect pair coverage by \avgDivImprove{}, which increases branch coverage by \avgCovImprove{}.
Further, we show that our context dependent custom mutation increases the proportion of valid tests by up to \maxContextValidImprove{}, 
indicating that \tool{} correctly concretizes its parameterized mutations with respect to the target context.
Parameterization of the mutations reduces the fraction of tests violating the base MLIR constraints by \paramGenMLIRReduction{}, increasing the time spent fuzzing dialect-specific code.
\end{abstract}

\begin{IEEEkeywords}
Grammar-based fuzzing, program synthesis, program transformation, MLIR, compiler testing, code patterns
\end{IEEEkeywords}

\section{Introduction}

Deep learning compilers are a critical component of AI workflows that enable PyTorch and TensorFlow models to be compiled to a variety of hardware architectures.
One of the leading technologies powering such DL compiler development is LLVM's Multi-Level Intermediate Representation (MLIR) framework.
Unlike LLVM, which defines a single common intermediate representation (IR), MLIR enables developers to extend the underlying IR through the concept of {\em MLIR dialects}~\cite{llvm-book, mlir-lang-ref}.
Each MLIR dialect defines a new IR consisting of a unique set of operations, types, and attributes with domain-specific semantics.
For example, the IR for machine learning is modeled a computation graph, while the IR for LLVM is modeled as a sequence of program instructions. 

Take the Circuit IR Compilers and Tools (CIRCT)~\cite{circt} project as an example. It leverages the MLIR framework to build a compiler for heterogeneous compilation by defining 26 new dialects with 145 new operations. For example, CIRCT uses a generic hardware abstraction by defining the \texttt{hw} and \texttt{comb} dialects with operations to represent abstract hardware modules and combinational logic. 
Listing \ref{lst:hw-example} shows a snippet of MLIR representing a hardware module containing a custom MLIR operation called \texttt{comb.add} which represents combinational addition.

Such fast evolution and diversity of underlying custom IRs presents challenges for developing custom test generators. Google reported in 2020 that over 60 dialects have been internally developed~\cite{google-mlir-tutorial}.
In the four years since the MLIR project's initial public release, 29 public downstream projects like CIRCT have also each contributed multiple custom dialects~\cite{circt,mlir-users}.
General-purpose fuzzers such as AFL++ \cite{AFLplusplus} fail to effectively generate or mutate MLIR due to its highly structured form.
For instance, syntactically correct MLIR must have proper nesting of operations, the correct number of operands and outputs for each operation, valid type annotations, and valid attribute names and values.
Grammar-based fuzzers, such as Grammarinator \cite{grammarinator}, use a context-free grammar to constrain input generation; however, there are two limitations for this domain. First, developers must supply a refined grammar that is specific to each MLIR dialect, which is a derivative of the base MLIR grammar. This refined grammar must include specialized production rules for each dialect's operation names, attribute names, and the number of inputs and outputs. For instance, to generate an \texttt{onnx.Conv2D} operation, the refined grammar would need a specialized production rule with operation name \texttt{onnx.Conv2D} that has 1 output, 3 inputs, and nested attribute names called \texttt{kernel\_size}, \texttt{padding}, and more. Creating such refined grammars by hand is impractical. For example, the four MLIR projects in our evaluation collectively define 74 dialects with 1,493 unique operations. Each operation would require at least one production rule in the refined grammar. Second, in addition to defining a refined grammar, developers must externally encode semantic constraints required by each custom dialect.

Custom generator-based fuzzers in the vein of CSmith \cite{csmith}, NNSmith \cite{nnsmith}, and MLIRSmith \cite{mlirsmith} manually encode semantic constraints in terms of imperative code. 
However, the large and continually increasing number of MLIR dialects makes it prohibitively expensive to write custom test generators manually.

We observe that the fast-evolving compiler infrastructure of IRs
requires a test input generator that can learn semantic constraints automatically. To this end, we propose a new approach called {\tool}, drawing inspiration from techniques for automated patch synthesis~\cite{meng2013lase,rolim2017learning,serrano2020spinfer}.
These techniques synthesize code edits from examples, 
learn the code contexts in which the transformations are appropriate and then concretize the code edits to the matching code contexts.

Like generator-based fuzzers, \tool{} is capable of preserving context-sensitive constraints such as
the cardinality of operation arguments and return values, the def-use relationships of values, and the consistency of type annotations.
The key novelty is that \tool{} can do so
without the significant manual effort required to write custom generators by hand.
\tool{} accomplishes this by synthesizing parameterized mutations from seed test cases.

\begin{figure}[t]
\begin{lstlisting}[
    label={lst:hw-example}, 
    caption={This MLIR code snippet uses a new hardware  dialect {\tt comb} for combinational logic in the CIRCT project.}, 
    language=mlir,
    escapeinside={|}{|}
]
"hw.module"() ({  // Hardware module definition
^bb0(|\hlcolor{hr}{\%a1}|: |\hlcolor{hp}{i2}|): 2-bit integer input
  // Defines a new constant with value -2, bitwidth 2:
  |\hlcolor{hg}{\%c1}| = "hw.constant"() {value = -2 : |\hlcolor{hp}{i2}|} : () -> |\hlcolor{hp}{i2}|
  // Adds the constants %0 and %1, outputting %2:
  |\hlcolor{hb}{\%o1}| = "comb.add"(|\hlcolor{hr}{\%a1}|, |\hlcolor{hg}{\%c1}|) <{twoState}> : (|\hlcolor{hp}{i2}|, |\hlcolor{hp}{i2}|) -> |\hlcolor{hp}{i2}|
  "hw.output"(|\hlcolor{hb}{\%o1}|) : (|\hlcolor{hp}{i2}|) -> () // Output of module is %2:
}) // extra boilerplate omitted
\end{lstlisting}
\vspace{-1.5em}
\end{figure}

Colored pairs in Listing~\ref{lst:hw-example} represent the def-use relationships and type consistency that needs to be satisfied. 
A parameterized mutation derived from Listing \ref{lst:hw-example} would encode the knowledge that the operation \texttt{comb.add} is nested within the \texttt{hw.module} denoted as ancestor $k_1$, is preceded by one \texttt{hw.constant} operation denoted as $l_1$, and followed by one \texttt{hw.output} operation denoted as $r_1$. This knowledge enables \tool{} to select an appropriate context to apply the mutation by matching the $k$-ancestors and $l$-siblings (and $r$-siblings) of the \texttt{comb.add} operation.
The parameterized mutation also encodes the knowledge that \texttt{comb.add} takes two arguments \texttt{\%a1} and \texttt{\%c1}, and returns one value \texttt{\%o1} all of which have the same type \texttt{i2}.
\tool{} parameterizes these arguments and types and then re-concretizes them based on the target context to which the mutation is applied.

We compare the effectiveness of \tool{} against Grammarinator, MLIRSmith, and NeuRI. 
Grammarinator is a representative grammar-based fuzzer.
MLIRsmith is a representative custom generator for MLIR core dialects.
NeuRI is a custom test generator with limited parameterization---i.e. it parameterizes the tensor shapes and operation's numerical attributes.
We evaluate \tool{} on four MLIR-based compiler projects: LLVM, ONNX-MLIR, Triton, and CIRCT. 
These are chosen as representative MLIR projects that define 42, 2, 4, and 26 custom dialects respectively.
For all dialects except the 13 core dialects targeted by MLIRSmith and the one \texttt{onnx} dialect that can be targetted by NeuRI, no custom test generators exist. 
Writing test generators for these custom dialects is time-consuming. 
As an example, MLIRSmith's implementation totals \MLIRSmithTotalLOC{} lines of code with \avgMLIRSmithLOCPerDialect{} lines of code per dialect on average~\cite{githubMLIRSmith}.

We assess \tool{}'s fault detection potential by measuring code coverage and MLIR dialect pair coverage.
Dialect pair coverage~\cite{mlirsmith} is defined as the number of unique pairs of operations/dialects that have a data dependency or control dependency.
Averaged across over four MLIR compiler projects, \tool{} outperforms Grammarinator, MLIRSmith, and NeuRI
in terms of branch coverage by \avgCovImproveGrammar{}, \avgCovImproveMLIRSmith{}, and \avgCovImproveNeuRI{}
respectively.
In terms of dialect pair coverage, \tool{} outperforms Grammarinator, MLIRSmith, and NeuRI on average by \avgDivImproveGrammar{}, \avgDivImproveMLIRSmith{}, and \avgDivImproveNeuRI{}.
Compared to MLIRSmith and NeuRI, \tool{} is capable of covering 60 new custom dialects defined by the four MLIR projects.
\tool{} discovers a previously undiscovered bug in CIRCT.

We also perform a case study on the potential of \tool{} to generalize beyond MLIR dialects to another domain by automatically generating valid AWS CloudFormation (CF) templates~\cite{cloudformation} that can pass the validity checks of \texttt{cfn-lint}~\cite{githubCfnLint}.
\tool{} generates 2.46$\times$ greater proportion of valid CF templates compared to Grammarinator, which demonstrates that \tool{}'s custom mutation synthesis can provide significant benefits to learn semantic constraints automatically.

In summary, this paper makes the following contributions:
\begin{enumerate}
    \item We design a novel compiler fuzzing technique that obviates the need for defining custom mutations apriori, which is impractical when the target IR is highly extensible and constantly evolving.  
    \item Our method automatically synthesizes and applies multi-edit, dependence-aware, custom mutations on the fly. The key enabler is the construction of parameterized mutations from existing tests, and the concretization of the mutations after positioning the context through ancestor path or prefix(postfix) matching.
    \item We show that our method achieves \avgCovImprove{} greater code coverage and \avgDivImprove{} greater dialect coverage within the same time budget compared to existing baseline fuzzers.
        \end{enumerate}

The remainder of this paper is organized as follows. 
Section~\ref{sec:motivating} introduces MLIR and a motivating example. 
Section \ref{sec:approach}  presents the design and implementation of \tool{}.
Section~\ref{sec:evaluation} provides the design of our experiments and their results.
Section~\ref{sec:discussion} discusses 
possible threats to validity.
Section~\ref{sec:related} presents related work.
Finally, we draw the conclusions of our work in Section~\ref{sec:conclusion}.

\section{Background}
\label{sec:motivating}
\subsection{MLIR: Multi-Level Intermediate Representation}

Multi-Level Intermediate Representation (MLIR) is a modular compiler framework that differs from traditional approaches by enabling developers to extend the intermediate representation.
Rather than defining a single monolithic IR with a fixed set of types and instructions like LLVM's IR, MLIR is extensible by design.
Compiler developers may define new \emph{MLIR dialects} consisting of custom operations and types tailored to the domain, language, or architecture the compiler targets. MLIR dialects can be progressively lowered, forming a modular compilation pipeline, in contrast with traditional compiler infrastructure that offers limited extensibility.
However, this presents a challenge for test generation, since dialect-specific operations also introduce new constraints that are dialect-specific. 
Take the Circuit IR Compilers and Tools (CIRCT)~\cite{circt} as an example.
CIRCT is a unified framework built on MLIR that enables optimized hardware design across different backends catering to the needs of heterogeneous compilation.
It defines 26 new dialects with 145 new operations, including the \texttt{comb} and  \texttt{hw} 
dialects that define low-level hardware operations. An example of the \texttt{comb} add operation is shown on line 4 of Listing \ref{lst:recipient-ex}. 
The full name of this operation is \texttt{comb.add} where \texttt{comb} is a dialect name and \texttt{add} is the operation name. 
As shown in the snippet, the operation takes two operands \texttt{\%arg0 and \%c1} and returns a single value \texttt{\%o1}. Its type signature indicates that the operation takes an input operands of type \texttt{i2} (2-bit integers) and produce an output of type \texttt{i2}.

\subsection{Motivating Example}

\begin{figure}

\begin{lstlisting}[
    label={lst:donor-ex}, 
    caption={A donor program $P_d$ from which a mutation for inserting the \texttt{comb.add} operation is synthesized from. 
        }
]
"hw.module"() ({
^bb0(|\textbf{\%arg0}|: |\textbf{i2}|):
  |\textbf{\%c1}| = "hw.constant"() {value = -2 : i2} : () -> |\textbf{i2}|
  |\textbf{\%o1}| = "comb.add"(|\textbf{\%arg0, \%c1}|) : (|\textbf{i2, i2}|) -> |\textbf{i2}|
  "hw.output"(|\textbf{\%o1}|) : (|\textbf{i2}|) -> ()
})
\end{lstlisting}
\begin{lstlisting}[
    label={lst:recipient-ex}, 
    caption={A recipient program $P_r$ to which the mutation for inserting the \texttt{comb.add} operation from Listing 2 should be applied to. }
]
"hw.module"() ({
^bb0(%arg0: i4, %arg1: !hw.array<2xi2>):
  |\textbf{\%0} = "hw.bitcast"(\textbf{\%arg1}) : (!hw.array<2xi2>) -> i4|
  |\textbf{\%1} = "comb.sub"(\textbf{\%arg0}, \textbf{\%0}) : (i4, i4) -> i4|
  |"hw.output"(\textbf{\%1}) : (i4) -> ()|
})
\end{lstlisting}

\begin{lstlisting}[
    label={lst:combine-basic}, 
    caption={A test case created by Grammarinator\cite{grammarinator}'s recombine operation. It deletes line4 and adds \hlcolor{PastelGreen}{line 5}. This test case is invalid as it violates the \hlcolor{PastelDark}{def-use} relation and the \hlcolor{PastelDark2}{type}} consistency.
]
"hw.module"() ({
^bb0(|\textbf{\%arg0}: \hlcolor{PastelDark2}{\textbf{i4}}, \%arg1: !hw.array<2xi2>):|
  |\textbf{\%0}| = "hw.bitcast"(|\textbf{\%arg1}|) : (!hw.array<2xi2>) -> i4
  |\sout{\textbf{\%1} = "comb.sub"(\textbf{\%arg0}, \textbf{\%0}) : (i4, i4) -> i4}|
  |\hlcolor{PastelGreen}{\textbf{\%o1} = "comb.add"(\textbf{\%arg0}, }\hlcolor{PastelDark}{\textbf{\%c1}}\hlcolor{PastelGreen}{) : (}\hlcolor{PastelDark2}{\textbf{i2}}\hlcolor{PastelGreen}{, \textbf{i2}) -> \textbf{i2}}|
  "hw.output"(|\textbf{\%1}|) : (i4) -> ()
})
\end{lstlisting}

\begin{lstlisting}[
    label={lst:combine-context}, 
    caption={A test case created by \tool{}'s context-dependent, parameterized mutation. This mutation replaces an operation \texttt{comb.sub} with \hlcolor{PastelGreen}{\texttt{comb.add}}. The test case is valid as \tool{} matched a corresponding context before \hlcolor{PastelBlue}{concretizing its mutation} to the target context. }, 
    escapeinside={|}{|}
]
"hw.module"() ({
^bb0(|\textbf{\%arg0}: \hlcolor{PastelBlue}{\textbf{i4}}, \%arg1: !hw.array<2xi2>):|
  |\textbf{\%0}| = "hw.bitcast"(|\textbf{\%arg1}|) : (!hw.array<2xi2>) -> i4
  |\sout{\textbf{\%1} = "comb.sub"(\textbf{\%arg0}, \textbf{\%0}) : (i4, i4) -> i4}|
  |\hlcolor{PastelBlue}{\textbf{\%1}}\hlcolor{PastelGreen}{ = "comb.add"(}\hlcolor{PastelBlue}{\textbf{\%arg0}, \textbf{\%0}}\hlcolor{PastelGreen}{) : (}\hlcolor{PastelBlue}{\textbf{i4}, \textbf{i4}) -> \textbf{i4}}|
  |"hw.output"(\hlcolor{PastelBlue}{\textbf{\%1}}) : (\hlcolor{PastelBlue}{\textbf{i4}}) -> ()|
})
\end{lstlisting}

\vspace{-1.5em}
\end{figure}

Existing mutation strategies such as recombining test fragments frequently fail to generate test cases capable of exercising deeper compiler logic. 
This failure is caused by the large proportion of \textit{invalid} test cases generated, which violate early checks made by the compiler.
For example, a) the definition of identifiers needs to exist before they are used (\textit{def-use}), b) the types of variables need to remain consistent through the test case (\textit{type consistency}), and c) the number and type of arguments that match what is required by an operation (\textit{signature consistency}). 
To address the limitation of existing fuzzers, 
we present an approach that synthesizes parameterized mutations, aiming to implicitly capture these constraints.

Consider the following seed test cases: a ``donor'' program $P_d$ in Listing \ref{lst:donor-ex} that contains the \texttt{comb.add} operation to be inserted in the ``recipient'' program $P_r$ in Listing \ref{lst:recipient-ex}. 
We demonstrate how grammar-based generation and recombination are unlikely to produce a valid program shown in Listing \ref{lst:combine-context}.

A grammar-based mutator following a base MLIR grammar is unlikely to produce Listing~\ref{lst:combine-context} because the grammar does not include operation semantics defined by different dialects.
For any given operation, the generic MLIR grammar only specifies that the syntax of an operation must have a name (e.g. \texttt{comb.sub}), zero or more return values, arguments and attributes, and a type signature.
Therefore, a grammar-based fuzzer generating a \texttt{comb.add} operation would be unaware of the signature of the operation as defined by the \texttt{comb} dialect.
Without a refined grammar for each dialect, the fuzzer would be unaware of the associations between variables and their types, e.g. \texttt{\%arg0}, \texttt{\%0} with type \texttt{i4}, and \texttt{comb.add} having exactly two input values and one return value.

Another common grammar-based mutation strategy is to recombine the \textit{fragments} of existing tests with other test cases.
To illustrate, \texttt{comb.sub} at line 4 of the recipient program $P_r$ in Listing \ref{lst:recipient-ex} is replaced with \texttt{comb.add} from line 4 of the donor program $P_d$ in Listing \ref{lst:donor-ex}, producing the mutated test in Listing \ref{lst:combine-basic}.
However, after the replacement, the values and types of the \texttt{comb.add} operation are inconsistent with its new surrounding context.
The resulting test will be rejected early by the compiler as it violates the def-use constraint, since the value \texttt{\%c1} was not defined before it was used.
Inferring such semantic constraints (often Turing-complete) is challenging as discussed in prior work~\cite{aschermann2019nautilus} \cite{wang2019superion} \cite{steinhofel2022input}.

Listing \ref{lst:combine-context} highlights the changes required to adapt the code using \texttt{comb.add} from  $P_d$ to $P_r$.
To satisfy the def-use constraint, 
the values referenced as arguments to an operation (e.g. \texttt{\%0} on line 5) must be previously defined (e.g. \texttt{\%0} on line 3). 
To satisfy type consistency, 
the initially assigned types, such as \texttt{i4}, for \texttt{\%0} on line 3 must remain consistent in its subsequent references, such as its use as an argument and the resulting return type on line 5.

To generate test cases that satisfy these constraints, developers either must write a custom generator or define new refined grammars and then manually encode additional semantic constraints.
However, this requires hand-coding the constraints of each operation defined in a dialect, and the cost is exacerbated for rapidly evolving projects with IR extensions---i.e. those that use MLIR---because the specialized grammar or custom generators will need to be updated as new dialect operations are added, modified, or removed.

In this paper, we propose to \textit{automatically synthesize custom mutators}. 
Existing test cases of MLIR dialects demonstrate how the dialect-specific operations should be invoked. Our key insight is that \textit{these test cases implicitly encode the various constraints of MLIR dialects, e.g., def-use, operations' type signature}, and therefore, matching the code context would lead to a higher chance of successfully generating valid inputs.  
For example, the donor test case in Listing~\ref{lst:donor-ex} and the recipient test case in Listing~\ref{lst:recipient-ex} exhibit structural similarity.
\tool{} is able to identify such similarity and synthesize a parameterized mutation that captures the context of the operation \texttt{comb.add} in the donor test case. Applying the mutation, \tool{} produces a valid test case as shown in Listing~\ref{lst:combine-context}, which respects dialect-specific constraints, and hence is able to exercise deeper logic in a given MLIR compiler.

\section{Approach}
\label{sec:approach}

\begin{figure}
    \centering
    \includegraphics[width=\linewidth]{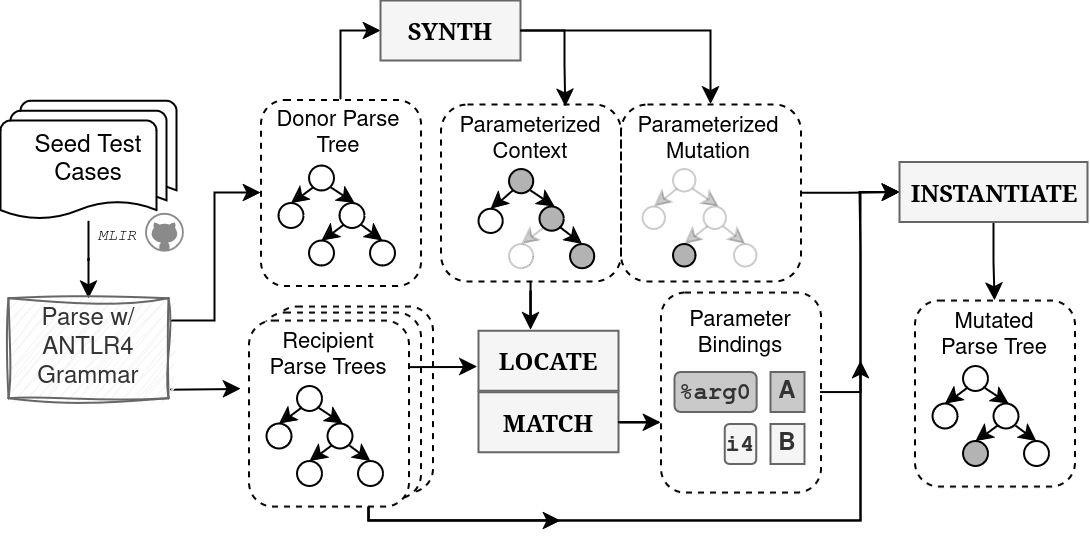}
    \caption{A flowchart of \tool{}'s fuzzing loop}
    \label{fig:synth-flowchart}
    \vspace{-1em}
\end{figure}

Figure~\ref{fig:synth-flowchart}
describes \tool{}'s fuzzing loop.
Each iteration synthesizes a custom mutation from a donor test case and transplants it onto a recipient test case. 
In the $\synth$ step 
(Section \ref{sec:synth}),
\tool{} selects a donor test case and a recipient test case.
From the donor test case, \tool{} infers a \textit{parameterized mutation} with a \textit{parameterized context}.
In the $\synthlocate$ step (Section \ref{sec:locate}), 
as the mutations are context-dependent, 
\tool{} matches the parametrerized context against the recipient test case to identify a suitable location for the mutation.
The $\synthmatch$ step (Section \ref{sec:match}) then creates a variable binding of parameters to concrete program fragments from the matching context.
Finally, $\instantiate$ (Section \ref{sec:instantiate}) concretizes the mutation and transplants it into the recipient test case.

\subsection{$\synth$: Synthesizing a parameterized mutation.}
\label{sec:synth}

\begin{figure}
    \centering
    \begin{minipage}{\linewidth}
        \begin{lstlisting}[label={lst:synth-example}, caption={In this donor test case $P_d$, the boxed area represents the source of a grafted, parameterized mutation. The rest unboxed area represents the potential source of a corresponding, parametrized context.}]
"hw.module"() ({
^bb0(|\hlcolor{hr}{\textbf{\%arg0}}|: |\hlcolor{hb}{\textbf{i2}}|):
  |\hlcolor{hp}{\textbf{\%c1}}| = "hw.constant"() {value = -2 : i2} : () -> |\hlcolor{hb}{\textbf{i2}}|
  |\fbox{\hlcolor{hg}{\textbf{\%o1}} = "comb.add"(\textbf{\hlcolor{hr}{\%arg0}, \hlcolor{hp}{\%c1}}) : (\hlcolor{hb}{\textbf{i2, i2}}) -> \hlcolor{hb}{\textbf{i2}}}|
  "hw.output"(|\hlcolor{hg}{\textbf{\%o1}}|) : (|\hlcolor{hb}{\textbf{i2}}|) -> ()
})
        \end{lstlisting}
        
    \end{minipage}
    \vspace{1.5em}     \begin{minipage}{\linewidth}
        \centering
        \includegraphics[width=0.9\linewidth]{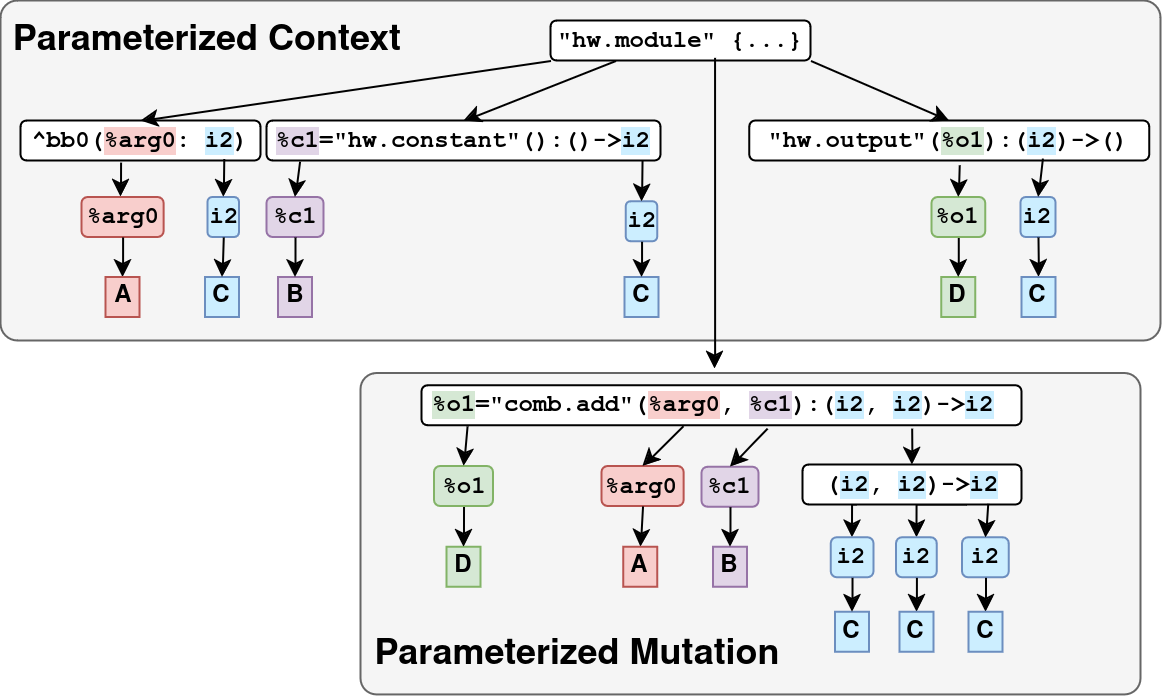}
    \end{minipage}
    \caption{This diagram illustrates how $\synth$ decomposes the donor test case $P_d$ shown in Listing~\ref{lst:synth-example} into a parameterized context and parameterized mutation. For example, concrete symbols such \texttt{\%arg0}, \texttt{\%c1}, \texttt{i2}, and \texttt{\%o1} are now paramterized as placeholders such as \hlcolor{hr}{A}, \hlcolor{hp}{B}, \hlcolor{hb}{C}, and \hlcolor{hg}{D} respectively.}
    \label{fig:synth-example}
\end{figure}

Parameterized mutations are synthesized from seed test cases which we will refer to as \textit{donor test cases}.
Given a donor test case like Listing~\ref{lst:synth-example}, \tool{} parses the test case using a base MLIR grammar to produce a \textit{donor parse tree}.
From the donor parse tree, \tool{} constructs a custom mutator comprising of a \textit{parameterized mutation} and a \textit{parameterized context} as shown in Figure~\ref{fig:synth-example}.

The parameterized mutation is a parameterized partial parse-tree that contains the content to be inserted by \tool{} when mutating a test case.   The parameterized context is similarly a partial parse tree that captures the conditions in which parameterized mutation may be instantiated.

$\synth$ uniformly randomly selects a sub-tree in the donor's parse tree as the parameterized mutation to be transplanted.
The parameterized context and parameterized mutation are extracted by bisecting the donor parse tree along at the selected sub-tree's root node as shown in Figure~\ref{fig:synth-example}.
The parameterized context encodes information such as the operations before and after the parameterized mutation, their ordering, the nesting of operations, and the potential locations of parameters.
For example, as shown in Figure~\ref{fig:synth-example}, the parameterized context encodes the information that \texttt{comb.add} is preceded by a block label \texttt{bb0} and the operation \texttt{hw.constant}, and succeeded by a \texttt{hw.output} operation.
It also encodes the nesting information that \texttt{bb0}, \texttt{hw.constant}, \texttt{comb.add}, and \texttt{hw.output} are nested within \texttt{hw.module}.

Since the donor's mutation may contain identifiers and types that are not defined in recipient programs, a naive transplantation will likely produce an invalid input that violates def-use and type consistency constraints. 
$\synth$ uses a heuristic that common sub-strings in the input can be indicators of context-dependent constraints such as def-use and type consistency.
$\synth$ parameterizes the context and mutation by introducing a parameter for each common sub-string between the context and the mutation.
In Figure~\ref{fig:synth-example} $\synth$ creates 4 parameters: A, B, C, and D with an initial binding of 
\texttt{\%arg0}, \texttt{\%c1}, \texttt{i2}, and \texttt{\%o1}
as illustrated with matching colors in Figure \ref{fig:synth-example}.
Later, each parameter can be concretized with a suitable sub-string from the context of the recipient.

\subsection{$\synthlocate$: Selecting mutation location depending on the target context }
\label{sec:locate}

\tool{} looks for locations where the parameterized context matches the recipient input to increase the likelihood of satisfying the constraints.
Three factors are considered: the number of matching ancestor nodes $k$, the number of left-sibling nodes $l$, and the number of right-sibling nodes $r$.
Matching $k$ ancestors with the parameterized context ensures that the mutation is made in a similar level of nesting.
This prevents situations such as improperly nested functions, or operations being inserted outside of functions.
Matching $l$ left-siblings and $r$ right-siblings with the parameterized context ensures that if an mutation pattern is of an operation, then the operation will likely be inserted in a location where the required number of operand values (left) is available, and the results of the parameterized mutation will be used accordingly (right).

\begin{algorithm}[ht]
\caption{$\synthlocate$: Finding a valid mutate location by matching $k$-ancestors and $l$($r$) siblings.} 
\label{alg:locate} 
\begin{algorithmic}[1]
\Require
    \item[]
        \begin{itemize}
            \item $context \gets$ the parametrized context
            \item $recipient \gets$ the recipient parse tree
        \end{itemize}
\Ensure
    \item[]
        \begin{itemize}
            \item $mutateLocation \gets$ a parse-tree node in the recipient test case that represents a valid mutate location
        \end{itemize}
\For{$candidate \gets \text{walk}(recipient)$}
    \State $isMatch \gets$ true
    \State \Comment{The following loop jointly assigns $\getNext$ and $m$}
    \ForB
        $(\getNext, m) \in \{\hspace{0.5em}(\getParent,k)$, \\
        \hspace{10.9em}$(\getLeftSibling,\hspace{1.1em}l)$, \\
        \hspace{10.9em}$(\getRightSibling,\hspace{0.5em}r)\}$
    \DoB
        \State $pNode \gets context$
        \State $cNode \gets candidate$
        \For{$i \in [0, m)$}
            \If{$\pName(pNode) \neq \pName(cNode)$}
                \State $isMatch$ $\gets$ false
                \State \textbf{break}
            \EndIf
            \State $pNode \gets \getNext(pNode)$
            \State $cNode \gets \getNext(cNode)$
        \EndFor
    \EndDoB
    \If{$isMatch$}
        \State \textbf{yield} candidate     \EndIf
\EndFor
\end{algorithmic}
\end{algorithm}

Algorithm \ref{alg:locate} describes how the mutate location is selected.
The values $k$, $l$, and $r$ are global hyper-parameters that are set by the user. In our evaluation, we set $k,l,r=4$.

On line 1, the $\text{walk}$ function returns each node of the recipient parse tree in breadth-first order. Each node considered a candidate mutate location.

On line 4 of the algorithm, $\getNext$ and $m$ are jointly assigned so that when $\getNext \gets \getParent$ then $m \gets k$ and when $\getNext \gets \getLeftSibling$ then $m \gets l$.
The $\getParent(n)$, $\getLeftSibling(n)$, and $\getRightSibling(n)$ functions on line 3 take a parse tree node $n$ and return the parent node of $n$, the node that precedes $n$ at the same depth, or the node that succeeds $n$ at the same depth respectively.
In Listing~\ref{lst:klr-recipient} calling $\getParent$, $\getLeftSibling$, and $\getRightSibling$ on \texttt{Location A} returns the enclosing block node (representing lines 2-7), the prior operation node (representing line 2), and the next operation node (representing line 5) respectively.

On line 11 of the algorithm, the $\pName(n)$ function returns the name of the production rule that corresponds with the parse-tree node $n$.
For example, calling $\pName$ on the \texttt{hw.bitcast} operation of Listing~\ref{lst:klr-recipient} returns the rule name "operation".

\begin{figure}[ht]
\centering
\begin{lstlisting}[
    label={lst:klr-recipient}, 
    caption={A set of potential insertion locations are marked as A, B, and C in the recipient test case.}, 
]
"hw.module"() ({
^bb0(%arg0: i4, %arg1: !hw.array<2xi2>):
  %0 = "hw.bitcast"(%arg1) : (!hw.array<2xi2>) -> i4
  |\hlcolor{hy}{\textbf{Location A}}|
  "hw.output"(%1) : (i4) -> ()
  |\hlcolor{hy}{\textbf{Location B}}|
}
|\;||\hlcolor{hy}{\textbf{Location C}}|
)
\end{lstlisting}
\includegraphics[width=0.8\linewidth]{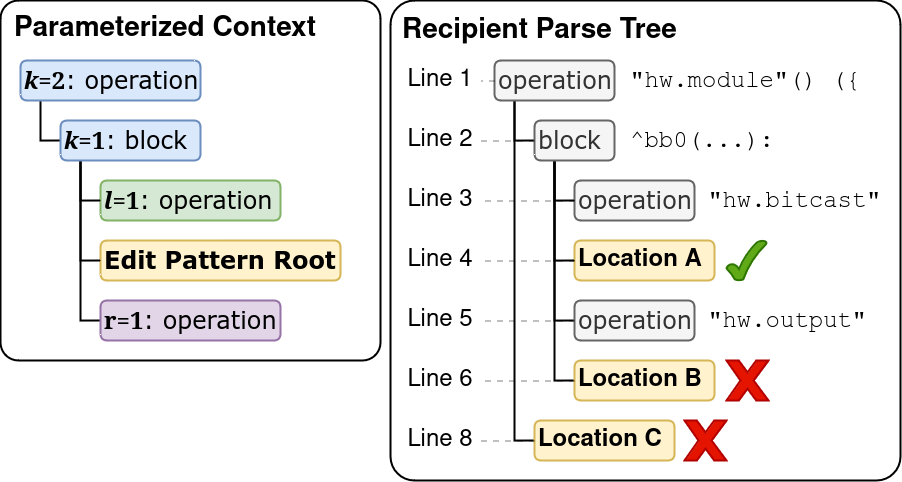}
\caption{Illustration of $k$-ancestor and $l(r)$-sibling context matching. Location B is invalid due to not matching the postfix context with $r=1$. Location C is invalid due to not matching the $k$-ancestor path context as the parent node is an operation, not a block with $k=2$.}
\label{fig:klr-match}
\end{figure}

Listings~\ref{lst:klr-recipient} and Figure~\ref{fig:klr-match} show an example of the $k$-ancestor, $l$-sibling, and $r$-sibling matching process.
$\synthlocate$ compares the parameterized context to each candidate location in the recipient test case.
Location A is a valid location since it has an operation as a left and right sibling (lines 3 and 5 respectively), and is nested within a block (line 2) and within a operation (line 1).
Location B is invalid because it is at the end of the block and thus has no right-siblings.
Location C is invalid because it has no right-siblings and its ancestor node is an operation, whereas the parameterized context requires the first ancestor to be a block.

Apart from replacing existing parse-tree nodes, \tool{} is also able to transplant content by inserting them in locations corresponding to quantifiers in production rules of the grammar.
Such quantifiers indicate that a varying length collection of terms can be generated.
\tool{} locates parse-tree nodes that correspond to production rule and inserts a new node corresponding to the quantified term.
\tool{} can then apply the same $k$-ancestor, $l(r)$-sibling matching logic described earlier to decide if the newly inserted node is a suitable mutation location.
This grants \tool{} greater flexibility, increasing the diversity of inputs generated.

\subsection{$\synthmatch$: Matching and extracting parameters}
\label{sec:match}

\begin{figure*}[t]
    \centering
    \includegraphics[width=0.8\linewidth]{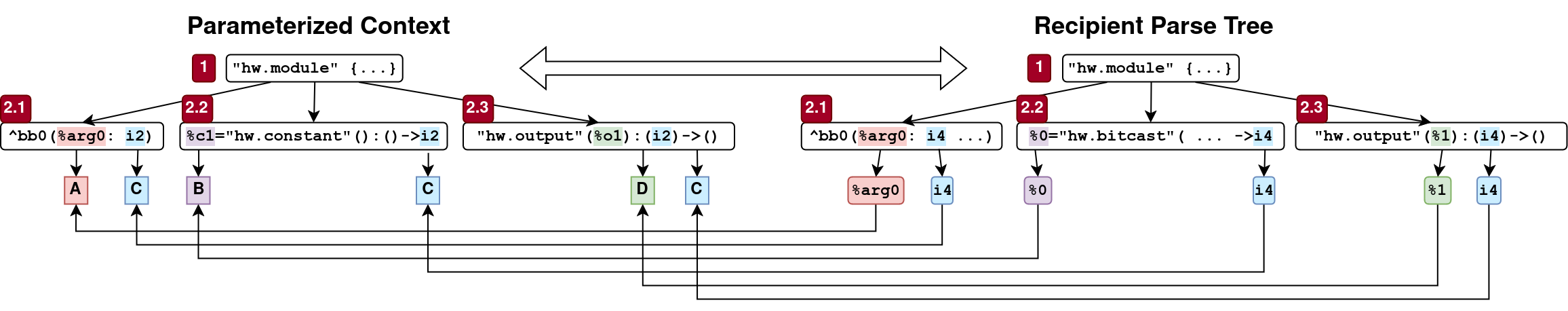}
    \caption{An illustration of the $\synthmatch$ step. When the recipient test case in Listing~\ref{lst:match-example} is matched with the parameterized context shown in this figure, the parameters \hlcolor{hr}{A}, \hlcolor{hp}{B}, \hlcolor{hb}{C}, and \hlcolor{hg}{D} are bound to the concrete values \texttt{\%arg0}, \texttt{\%0}, \texttt{\%i4}, and \texttt{\%1} respectively.}
    \label{fig:match-example}
\end{figure*}

Once an mutation location is selected in the $\synthlocate$ step, $\synthmatch$ performs a joint breadth-first traversal of the parameterized context tree and the parse-tree of the recipient test case to compute parameter assignments.
The traversal starts at the mutate location of the recipient tree and the root of the parameterized mutation within the parameterized context.
An annotated example of this traversal is illustrated in Figure~\ref{fig:match-example}.
A node pair is considered a match if the parse tree node name is the same between the parameterized context and recipient parse tree.
This allows parse tree nodes of different concrete operations to match.
For example, node pair 2.2 in Figure~\ref{fig:match-example} is considered a match even though the operation names (\texttt{hw.constant} and \texttt{hw.bitcast}), types (\texttt{i2} and \texttt{i4}), and values (\texttt{\%c1} and \texttt{\%0}) differ.

When a parameter node is encountered during the traversal of the parameterized context, \tool{} assigns it to the corresponding sub-tree in the traversal of the parse tree of the recipient input.
In Figure~\ref{fig:match-example}, the parameters A, B, C, and D from the parameterized context are assigned to the matching nodes for \texttt{\%arg0}, \texttt{\%0}, \texttt{i4}, and \texttt{\%1} in the recipient parse tree.
When there are duplicate assignments as in parameter C, $\synthmatch$ will uniformly randomly select one of the assignments to use.

\subsection{$\instantiate$: Concretize the mutation}
\label{sec:instantiate}

\begin{lstlisting}[label={lst:match-example}, caption={The recipient test case where the parametrerized mutation is inserted and concretized. The parametrerized mutation is boxed on line 4. By instantiating the pattern in this new context, the following substitutions are made: \hlcolor{hr}{A}$\gets$\texttt{\%arg0}, \hlcolor{hp}{B}$\gets$\texttt{\%0}, \hlcolor{hb}{C}$\gets$\texttt{\%i4}, and \hlcolor{hg}{D}$\gets$\texttt{\%1}.
}]
"hw.module"() ({
^bb0(|\hlcolor{hr}{\textbf{\%arg0}}: \hlcolor{hb}{\textbf{i4}}, \%arg1|: !hw.array<2xi2>):
  |\hlcolor{hp}{\textbf{\%0}} = "hw.bitcast"(\%arg1) : (!hw.array<2xi2>) -> \hlcolor{hb}{\textbf{i4}}|
  |\fbox{\hlcolor{hg}{\textbf{\%1}} = "comb.add"(\textbf{\hlcolor{hr}{\%arg0}, \hlcolor{hp}{\%0}}) : (\hlcolor{hb}{\textbf{i4, i4}}) -> \hlcolor{hb}{\textbf{i4}}}|
  |"hw.output"(\hlcolor{hg}{\textbf{\%1}}) : (\hlcolor{hb}{\textbf{i4}}) -> ()|
})
\end{lstlisting}

In the $\instantiate$ step, \tool{} adapts the parameterized mutation to the recipient test case by substituting in the parameter assignments extracted during the $\synthmatch$ step. Listing \ref{lst:match-example} shows how the parametrerized mutation is instantiated.
Here, parameters A, B, C, and D corresponding the return value, two operands, and the types are assigned the values 
\texttt{\%arg0}, \texttt{\%0}, \texttt{\%i4}, and \texttt{\%1} respectively.

\tool{} also checks that the mutated input conforms to generic MLIR constraints before passing them to the compiler.
For example, a mutation may lead to test cases that redefine the same variable twice or use an undefined variable. 
Such test cases are invalid in any MLIR dialect.

\section{Evaluation}
\label{sec:evaluation}
In our study, we examine the following research questions:
\begin{enumerate}[label=\textbf{RQ\arabic*:}]
    \item \label{rq:coverage} How effective is \tool{} in terms of increasing code coverage?
    \item \label{rq:diversity} What is the input space coverage of \tool{} in terms of dialect pair coverage?
    \item \label{rq:ablation-context} Does context-based positioning of parametrized mutation improve the likelihood of a valid mutation?
    \item \label{rq:ablation-param} Does parameterization of the mutation content improve the likelihood of a valid mutation?
\end{enumerate}

\subsection{Experiment Design}
\vspace{-2em}
\begin{table}[h]
    \caption{Benchmark programs}      \label{tab:subjects}
    \begin{tabular}{L{5em}L{11em}R{5em}R{4em}}
      \toprule
      \textbf{Subject Program} & \textbf{Description}  & \textbf{\# of Seed \mbox{Test Cases}} & \textbf{\# of Dialects} \\
      \midrule
      mlir-opt (P1) & Includes the core and contributed MLIR dialects part of the LLVM project. & 1,692 & 42 \\
      \midrule
            onnx-mlir-opt (P2) & An MLIR-based ONNX compiler. & 1,885 & 2 \\
      \midrule
      triton-opt (P3) & An MLIR-based compiler for the Triton language. & 29 & 4 \\
      \midrule
      circt-opt (P4) & An MLIR-based compiler for electronic design automation (EDA). & 377 & 26 \\
      \bottomrule
  \end{tabular}
  \vspace{0.5em}

  {Each project contributes a number of custom dialects for which it would be costly to hand-write test generators.}
\end{table}

In our evaluation of \tool{}, we select four active projects that use the MLIR compiler infrastructure, shown in Table \ref{tab:subjects}.
Each project provides a utility executable named \texttt{<project>-opt} which is used to independently invoke and test one or more compiler passes.
Although \tool{} can be used to test compilers end-to-end by generating MLIR inputs consumed by the frontend of the compilers, we found it more effective to generate MLIR inputs that can be directly consumed by intermediate compiler passes.
To fuzz each project, we invoke the \texttt{<project>-opt} executable with a pipeline of $P$ randomly selected compiler passes on each test case generated by a fuzzer.
In practice, we set $P=5$ since most unit test cases written by developers only invoke one to three compiler passes at a time.

For ONNX-MLIR, LLVM, and CIRCT, we select compiler passes based on the dialects present, identified by the operator names in the test case.
For example, if a test case contains the \texttt{arith.maxsi} operation, then our test driver will select a compiler pass that operates on the \texttt{arith} dialect such as the \texttt{arith-to-llvm} pass which lowers certain \texttt{arith} operations to the \texttt{llvm} dialect. 
As Triton's compiler passes do not follow this naming scheme and this heuristic cannot be used, we select compiler passes randomly from all available passes.

\subsubsection{Locating Seed Test Cases}
We build a corpus of seed test cases for each subject project by locating and splitting ``.mlir'' files in each subject project's respective  repositories for a total of 1,692, 1,885, 26, and 377 seed test cases respectively. Then we convert each test case into its generic MLIR syntax form to remove syntactic sugar and enable it to be parsed using the base MLIR grammar for all subject programs.

\begin{lstlisting}[label={lst:control-dep-ex}, caption={\texttt{hw.constant} (line 3) and \texttt{comb.icmp} (line 4) have a control dependence on \texttt{sv.if} (line 2), forming two dependent dialect pairs (\texttt{comb}, \texttt{sv}) and (\texttt{hw, sv}). \texttt{comb.icmp} (line 4) also has a data dependence on \texttt{hw.constant} (line 3) forming a data dependent dialect pair (\texttt{scf}, \texttt{hw}).}]
  |\textbf{"sv.if"}(\%arg2)| ({
    |\textbf{\%18}| = |\textbf{"hw.constant"}|() {value = 10 : i32} : () -> i32
    |\%19 = \textbf{"comb.icmp"}(\%7, \textbf{\%18}|) <...> : (i32, i32) -> i1
\end{lstlisting}

\subsubsection{Evaluation Measures}
\label{sec:eval-measures}
To compare the fault detection potential of \tool{}, we use the following measures:
\begin{itemize}
    \item \textbf{Branch coverage} is computed as the number of covered branches as reported by LLVM's SanitizerCoverage code coverage instrumentation.     
    \item \textbf{Dialect coverage} is a measure of input space coverage used in the evaluation of prior work, MLIRSmith~\cite{mlirsmith}. This notion can be further divided into control dependent dialect coverage and data dependent dialect coverage. The idea is to differentiate the usage of multiple dialects in tandem, and to prefer the usage of distinct dialects in a meaningful manner by being connected through data or control dependencies.
    \begin{itemize}
        \item \textbf{Control dialect coverage} is computed by counting the number of unique dialect pairs whose operations are linked by a control dependence. 
        For example, Listing~\ref{lst:control-dep-ex} contains two control dependent dialect pairs, (\texttt{comb}, \texttt{sv}) and (\texttt{hw, sv}), since \texttt{hw.constant} (line 2) and \texttt{comb.icmp} (line 3) are nested within \texttt{sv.if} (line 1) forming a control dependence.
        \item \textbf{Data dialect coverage} is computed by counting the number of unique dialect pairs whose operations are linked by a data dependency.
        For example, Listing~\ref{lst:control-dep-ex} contains one data dependent dialect pair, 
        (\texttt{scf}, \texttt{hw}) since 
        \texttt{comb.icmp} (line 3) takes the output \texttt{\%18} of \texttt{hw.constant} (line 2), thus forming a data dependence.
    \end{itemize}
\end{itemize}

\subsubsection{Baselines}
\begin{figure*}[ht]
    \centering
    \includegraphics[width=0.8\linewidth]{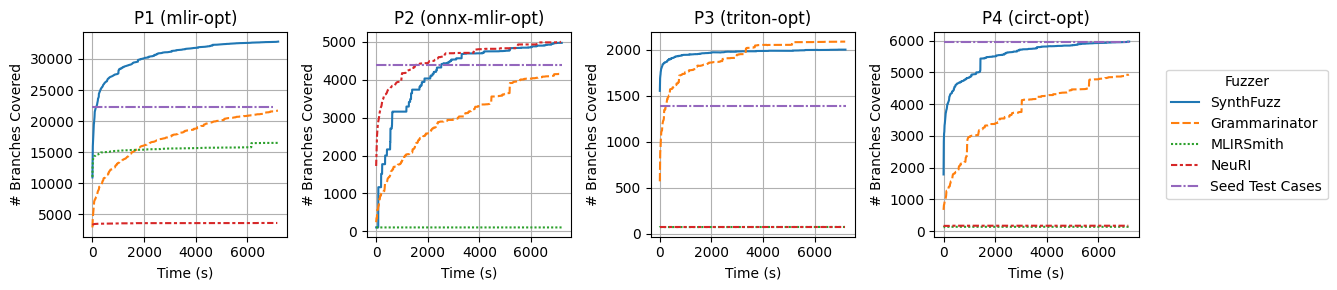}
    \caption{Branch coverage for each subject program.  \tool{} outperforms a grammar-based fuzzer by up to 1.51$\times$ and improves coverage by up to 1.47$\times$ compared to existing seed tests. }
    \label{fig:branch-coverage}
\end{figure*}

We evaluate \tool{} against Grammarinator~\cite{grammarinator}, MLIRSmith~\cite{mlirsmith}, and NeuRI~\cite{neuri}.
Grammarinator represents a baseline grammar-based fuzzer, while MLIRSmith and NeuRI represent the state of the art custom generators for MLIR and deep learning models. 

Grammarinator converts ANTLR4 grammar definitions into a random probabilistic generator code that can be used to generate, mutate, and recombine test cases based on the grammar rules. In our experiments, we provide both \tool{} and Grammarinator with the same seed corpus for a fair comparison.

MLIRSmith is a custom generator-based fuzzer that targets MLIR's core dialects.
MLIRSmith was chosen for comparison as it specializes in generating valid MLIR programs for MLIR's core dialects, but requires users to implement custom generators for other new dialects. As of March 2024, MLIRSmith supports 13 out of 42 available core dialects~\cite{githubMLIRSmith}.
NeuRI is a DL model-level fuzzer that generates computation graphs based on DL-model specific, API-level constraints.
NeuRI supports parameterization in the form of tensor shape constraints and shape propagation rules for machine learning models.
While NeuRI does not directly target MLIR, the \texttt{onnx-mlir} tool can be used to lower the models generated by NeuRI to the core MLIR dialects for P1 (mlir-opt) or to the \texttt{onnx} dialect for P2 (onnx-mlir-opt).

We also compare \tool{} against running existing test cases (i.e., the seeds for \tool{} and Grammarinator). By measuring the branch and dialect coverage of the seed test cases alone, we separate the contributions of the increased coverage afforded by the fuzzers' mutation capability from the innate capability of the seed corpus they draw from.

\subsubsection{Experimental Environment}
All experiments are performed on an AMD Ryzen 2950X 16-Core Processor with 32 GB of RAM running on Ubuntu 22.04.

\subsection{\ref{rq:coverage} Branch Coverage}

Figure \ref{fig:branch-coverage} shows the branch coverage of \tool{} and the baseline fuzzers on the four subject programs. 
Averaging across all subject programs, \tool{} outperforms Grammarinator, MLIRSmith, and NeurRI by \avgCovImproveGrammar{}, \avgCovImproveMLIRSmith{}, and \avgCovImproveNeuRI{} respectively.

On P1 (mlir-opt) \tool{} outperforms Grammarinator, MLIRSmith, and NeuRI by \avgCovImproveGrammarMLIR{}, \avgCovImproveMLIRSmithMLIR{}, and \avgCovImproveNeuRIMLIR{} respectively. 
On P4 (circt-opt) \tool{} outperforms Grammarinator, MLIRSmith, and NeuRI by \avgCovImproveGrammarCIRCT{}, \avgCovImproveMLIRSmithCIRCT{}, and \avgCovImproveNeuRICIRCT{} respectively. 

On P2 (onnx-mlir-opt), \tool{} is similar to NeuRI, with less than 1\% difference in coverage.  This demonstrates that \tool{} can match the performance of a domain-specific fuzzer, NeuRI, by bootstrapping parameterized mutations from existing test cases without hand-coding any custom generator logic. NeuRI implements custom test generator logic for ONNX models (computation graphs for DL models).

On P3 (triton-opt), Grammarinator outperforms \tool{} in terms of branch coverage by 1.04$\times$.
P3 only provides 36 test cases in its repository that \tool{} and Grammarinator could use as seeds.
Since \tool{} relies upon seed test cases to synthesize its custom mutations,
the low number of seeds constrained \tool{}'s ability to generate diverse test cases.

\textbf{An example bug found.}
\tool{} discovered a new bug in CIRCT (Issue \#6799), which has been confirmed and fixed by the developers.
This bug occurs when the \texttt{--convert-llhd-to-llvm} pass is invoked on an \texttt{llhd.proc} operation with a block without a terminator operation, such as \texttt{llhd.wait} or \texttt{llhd.halt}.
The CIRCT compiler incorrectly assumes that the \texttt{llhd.proc} operation always contains a terminator, and crashes due to the violation of this dialect-specific requirement. 
The CIRCT verifier should have detected this absence of a terminator and rejected the \texttt{llhd.proc} operation.

\begin{lstlisting}[label={lst:circt-bug}, caption={CIRCT crashes on this minimized input due to not checking for a terminator within \texttt{llhd.proc}. Reported on Mar 7, 2024, the bug was fixed immediately \mbox{on Mar 8.}}, numbers=none]
module {llhd.proc @empty() -> () { }}
\end{lstlisting}

\begin{resultbox}
\tool{} demonstrates an average \avgCovImprove{} improvement of branch coverage without requiring any hand-coding of custom generator logic.
\end{resultbox}
\subsection{\ref{rq:diversity} Dialect Coverage}
\begin{table*}[ht]
    \centering
    \caption{Dialect coverage for each subject program.}
    \label{tab:dialect-pair-cov}
\setlength\tabcolsep{5pt} \begin{tabular}{l|rrr|rrr|rrr|rrr}
\toprule
Subject & \multicolumn{3}{r|}{P1 (mlir-opt)} & \multicolumn{3}{r|}{P2 (onnx-mlir-opt)} & \multicolumn{3}{r|}{P3 (triton-opt)} & \multicolumn{3}{r}{P4 (circt-opt)} \\
 & dialects & control & data &  dialects & control  & data  &  dialects & control  & data  &  dialects & control  & data  \\
\midrule
SynthFuzz & 27 & \textbf{129} & \textbf{100} & \textbf{9} & \textbf{23} & \textbf{24} & \textbf{7} & \textbf{15} & \textbf{4} & 20 & \textbf{86} & \textbf{40} \\
\midrule
Seed tests as is& \textbf{28} & 68 & 56 & \textbf{9} & 17 & 11 & \textbf{7} & 12 & \textbf{4} & \textbf{23} & 57 & \textbf{40} \\
Grammarinator & 27 & 63 & 49 & 8 & 15 & 8 & \textbf{7 }& 12 & \textbf{4} & 17 & 43 & 27 \\
MLIRSmith & 13 & 62 & 65 & 4 & 6 & 3 & 3 & 3 & 1 & 6 & 15 & 10 \\
NeuRI & 7 & 17 & 12 & 6 & 9 & 10 & 3 & 3 & 2 & 6 & 14 & 10 \\
\bottomrule
\end{tabular}
\vspace{0.5em}

{Seed test cases refers to each subject's respective test suite. \tool{} achieves greater dialect coverage compared to baseline fuzzers.}
\end{table*}

Table~\ref{tab:dialect-pair-cov} summarizes the performance of \tool{} as compared to Grammarinator, MLIRSmith and NeuRI in terms of dialect control/data pair coverage (described in Section \ref{sec:eval-measures}).
SynthFuzz outperforms Grammarinator, MLIRSmith and NeuRI by \avgConDivImproveGrammar{} and \avgConDivImproveMLIRSmith{} and \avgConDivImproveNeuRI{} in terms of control dependent dialect pairs and \avgDataDivImproveGrammar{}, \avgDataDivImproveMLIRSmith{}, \avgDataDivImproveNeuRI{} in terms of data dependent dialect pairs.

To validate whether \tool{} discovers new dialect pairs not covered by the seed corpus, we also measure the dialect pair coverage of the seed corpus as a baseline. 
For each subject and fuzzer combination, we report the number of unique dialects and the number of unique control and data dialect pairs.
Across all subjects, \tool{} covers \newConPairs{} new control-dependent and \newDataPairs{} new data-dependent dialect pairs that did not already exist in the seed corpus.

\begin{lstlisting}[label={lst:synthfuzz_unique}, caption={A test case generated by \tool{} that was not generated by any baselines. 
\tool{} inserts the underlined line 5 and substitutes in the operand \texttt{\%1} and type \texttt{i32}, satisfying the required def-use and type consistency constraints.
This test case achieves a new dialect pair coverage, (\texttt{comb}, \texttt{arith}) by introducing the \texttt{comb.icmp} operator.  }]
"func.func"() <{function_type = (i1, i32, i32) -> i32, sym_name = "main"}> ({
^bb0(%arg0: i1, %arg1: i32, %arg2: i32):
  %0 = "arith.addi"(%arg1, %arg2) <{overflowFlags = #arith.overflow<none>}> : (i32, i32) -> i32
  %1 = "arith.shli"(%0, %arg1) <{overflowFlags = #arith.overflow<none>}> : (i32, i32) -> i32
  |\uline{\%6 = "comb.icmp"(\%1, \%1) <{predicate = 1 : i64}> : (i32, i32) -> i1}|
  %2 = "arith.subi"(%1, %0) <{overflowFlags = #arith.overflow<none>}> : (i32, i32) -> i32
  %3 = "arith.select"(%arg0, %3, %0) : (i1, i32, i32) -> i32
  "func.return"(%4) : (i32) -> ()
}) : () -> ()

\end{lstlisting}

Listing \ref{lst:synthfuzz_unique} shows an example test case that can be generated by \tool{}, but cannot be generated by other approaches.
The \texttt{comb} dialect is not supported by MLIRSmith, as it currently implements test generator logic for 13 core dialects only and takes \avgMLIRSmithLOCPerDialect{} lines of code on average to support each additional dialect.
Grammarinator's naive recombination fails to satisfy def-use and type consistency constraints by inserting line 5 as \texttt{\%6 = "comb.icmp"(\%5, \%4) <{predicate = 0 : i64}> : (i2, i2) -> i1 }, which references an undefined variable \texttt{\%5} and uses an incorrect type \texttt{i2} for variable \texttt{\%4}. This is because Grammarinator's recombination is context-unaware and is not concretized to fit the target context.   

While Grammarinator takes the same set of seed tests as \tool{}, it achieves  lower dialect coverage than \tool{} or the seed tests alone, because Grammarinator alternates between generation, mutation, and recombination modes. Approximately, one third of its time is spent on the pure generation mode that does not use seed test cases. 

To investigate \tool{}'s capability to generate test cases tailored for each dialect, we compute the proportion dialect-specific operations (e.g. \texttt{onnx.Add}) to total operations (e.g. including non-interesting operations like \texttt{func.return}) in the generated inputs. We find the proportions to be 0.49, 0.48, 0.92, and 0.87 for test cases from LLVM, ONNX-MLIR, Triton, and CIRCT respectively. This shows \tool{}'s potential to generate test cases with a high proportion of dialect-specific operations by synthesizing mutations from existing tests.

\begin{resultbox}
 \tool{}  achieves greater dialect  coverage, compared to
other baseline fuzzers: average \avgConDivImprove{} improvement in terms of control-dependent dialect pairs and average \avgDataDivImprove{} in data-dependent dialect pairs.
\end{resultbox}
\subsection{\ref{rq:ablation-context} Context-based Positioning of  Mutation}

\tool{} selects an appropriate insertion location to inject a parameterized mutation by matching the mutation's parameterized context against the target context of the recipient test case. To test the individual effect of the context matching requirement for a parameterized mutation, we vary $k$ from 0, 2, and 4 when matching a $k$-ancestor path. Similarly, we vary $l$ from 0, 2 and 4, when matching a $l$-sibling prefix $l$, and we vary $r$ from 0, 2, and 4 when matching a $r$-sibling postfix. Each trial consists of 10,000 test cases generated by \tool{} for P1 (mlir-opt). 

As shown in Table~\ref{tab:taguchi-response}, setting each parameter $k$, $l$, and $r$ to 4 improves the number of valid test cases by \kFourValidImprove{}, \lFourValidImprove{}, and \rFourValidImprove{} respectively. This indicates that using more context information increases the chance of finding an appropriate location for injecting a grafted mutation, thus increasing the portion of valid test cases. A test case is considered \textit{valid}, if feeding the generated input to the target program returns zero, indicating a success.

Setting $k=4$ decreases dialect pair coverage by 10\%. This may indicate that a very restrictive requirement for matching context by increasing $k$ can negatively affect the input coverage of generated tests. Additional experiments with k greater than 4 had minimal effect on the number of valid tests, as most seed tests have an operation nesting depth less than 4.

\begin{table}
\centering
\caption{Average branch coverage, dialect pair coverage, and valid test cases.}
\label{tab:taguchi-response}
\begin{tabular}{l|rrr}
\toprule
Parameter & Branch Cov. & Dialect Pair Cov. & Valid Test Cases \\
\midrule
$k=0$ & 24,764 & \textbf{100} & 677 \\
$k=2$ & 24,765 & \textbf{100} & 698 \\
$k=4$ & \textbf{24,987} & 90  & \textbf{749} \\
\midrule
$l=0$ & \textbf{25,055} & 94  & 684 \\
$l=2$ & 24,749 & 97  & 710 \\
$l=4$ & 24,713 & \textbf{98}  & \textbf{729} \\
\midrule
$r=0$ & 24,657 & 97  & 706 \\
$r=2$ & 24,901 & 95  & 687 \\
$r=4$ & \textbf{24,958} & \textbf{98}  & \textbf{731} \\
\bottomrule
\end{tabular}
\vspace{0.5em}

{Increasing $k$, $l$, and $r$ to 4 increases the number of valid test cases by 1.11$\times$, 1.07$\times$, and 1.03$\times$}
\end{table}

\begin{resultbox}
Increasing ancestor-path, prefix, and postfix requirements for context positioning improves the proportion of valid test cases.
\end{resultbox}

\subsection{\ref{rq:ablation-param} Effect of Parameterization}

\tool{} has the capability to parameterize and concretize a variable name, an argument's type, and an operation's attribute (e.g. \texttt{"hw.constant"() \{value = -2 : i2\}} has a value attribute \texttt{-2} with the type \texttt{i2}) to fit the target context where a grafted mutation is inserted into. 
The goal of parameterization is to preserve context-sensitive constraints such as the def-use constraint (i.e., a value must be defined before use) and the type-constraint (i.e., the type annotation of a value must be consistent throughout its scope).

\begin{table}[h]
    \centering
    \caption{Validity of test cases generated by \tool{}.} 
    \label{tab:parameter-ablation}
\begin{tabular}{lllrr}
\toprule
 & Violation Type &  & W/ Param. & W/O Param. \\
\midrule
\multirow[t]{6}{*}{Invalid} & \multirow[t]{2}{*}{Dialect Specific} & Count & 4,259 & 2,450 \\
 &  & Percent & 38.1\% & 23.7\% \\
\cmidrule{2-5}
 & \multirow[t]{2}{*}{General MLIR} & Count & 1,777 & 3,120 \\
 &  & Percent & 15.9\% & 30.2\% \\
\cmidrule{2-5}
 & \multirow[t]{2}{*}{Invalid Options} & Count & 4,356 & 4,052 \\
 &  & Percent & 39.0\% & 39.2\% \\
\cmidrule{1-5}
\multirow[t]{2}{*}{Valid} & \multirow[t]{2}{*}{Valid} & Count & 772 & 702 \\
 &  & Percent & 6.9\% & 6.8\% \\
\bottomrule
\end{tabular}
\vspace{0.5em}

{Parameterization reduces General MLIR violations from 30.2\% to 15.9\%.}
\end{table}

We create a downgraded version of \tool{} by turning off its parameterization and concretization capability denoted as W/O Param in Table~\ref{tab:parameter-ablation}.
We generate 10,000 test cases with each version and categorize the test cases based on the error message returned by the target program.

Parameterization increases the proportion of valid test cases by 0.01\% only. 
However, when we further inspect the underlying reasons for invalid test cases, we find that \tool{} increases the chance of adhering to the general MLIR constraints.

With parameterization, 772 tests are valid with the return value zero indicating success, when the tests are fed to the target program. We then categorize the remaining 9228 invalid test cases into three categories based on the type of violation reported by the target program. 
\begin{itemize}
    \item \textbf{Dialect Specific}: 4,259 test cases generated with parameterization are rejected by the target program with a dialect-specific error message such as: \texttt{tosa.logical\_or} op result \#0 must be tensor of 1-bit signless integer values, \texttt{tosa.floor} op requires a single operand, etc.
    \item \textbf{General MLIR}: 1,777 test cases generated with parameterization are rejected by the target program with a general MLIR error message such as: {\em an undefined symbol}, {\em use of undeclared SSA}, {\em redefinition of SSA value}, etc.
    \item \textbf{Invalid Options}: 4,356 test cases generated with parameterization are rejected with an error message, ``no such option exists.'' This occurs due to the test driver's random pass selection which may pair an option with a pass that does not accept said option.
\end{itemize}

With parameterization enabled, \tool{} generates 1,343 fewer test cases that violate general MLIR constraints out of 10,000 tests. 
The proportion of Invalid Options category is approximately the same with and without parameterization. However, the proportion of General MLIR invalidity increases from 15.9\% to 30.2\% when disabling parameterization.
This is due to the fact that without parameterization, the content of parameterized mutation is not concretized to fit the recipient context. Thus it is more likely to violate general MLIR constraints such as def-use and type consistency.

Listing~\ref{lst:synthfuzz_unique} shows an example of a test case generated by \tool{} which nests the \texttt{comb.icmp} operation within a \texttt{func.func} operation.
\tool{} parameterizes  the input arguments and their types, thus passing the general MLIR constraints such as def-use and type consistency.

\begin{resultbox}
\tool{} reduces the proportion of general MLIR constraint violating tests from 30.2\% to 15.9\% by parameterizing the injected mutation's content.
\end{resultbox}

\section{Discussion}
\label{sec:discussion}

\subsection{Case study using a grammar-aware constraint solver (ISLa)}
\label{sec:isla}

ISLa is a grammar-based constraint solver that can act as a fuzzer to generate inputs~\cite{steinhofel2022input}. ISLa enables a user to specify SMT-like constraints on top of a context-free grammar. It then generates inputs to satisfy grammar-aware string constraints. To fuzz MLIR-based compilers, ISLa requires a user to define a precise, refined grammar and manually encode input constraints for each MLIR dialect. On the other hand, \tool{} infers implicit constraints specific to each dialect by leveraging existing tests.

As a case study, we used ISLa under two configurations on the same four subject programs: Under configuration (A), we provided ISLa the same base MLIR grammar provided to Grammarinator and \tool{}. This base MLIR grammar is generic in that it accepts all MLIR dialects. However, it is not specialized to any MLIR dialect and thus it does not encode any dialect specific operations, types, or attributes. Under configuration (B), we provided ISLa with a specialized refined grammar for a subset of the ONNX MLIR dialects, consisting of 25 custom ONNX MLIR operations.

Under both configurations, we used ISLa's input specification language to encode two constraints: one definition-before-use and one no-redefinition.
For configuration (B), we  encoded a third constraint, a simplified type-consistency constraint.

\subsubsection{Results}
We ran ISLa under configurations (A) and (B) with the same four hour time budget used in our evaluation. We found that when ISLa is configured to generate more than 10 inputs, the solver frequently stalls indefinitely, likely due to the complexity of the generated SMT constraints. To work around this issue, we restarted ISLa every 10 inputs to maintain an approximate throughput of 3 seconds per input.

In both configurations, ISLa did not generate any valid MLIR inputs. Inputs generated by configuration (A) were invalid because none of the operation names are specific to an individual MLIR dialect. In configuration (B), the refined grammar specialized to ONNX MLIR along with the three constraints caused ISLa to generate only MLIR inputs with the \texttt{NoValue} operation (equivalent to a no-op). Thus, we had to disable the ISLa constraints to generate the remaining 24 operations defined in our refined grammar. In such case, all inputs failed the ONNX MLIR compiler's input validity checks as they violated the def-use, no-redef, and type-consistency constraints. Removing the \texttt{NoValue} operation from the grammar  caused the ISLa solver to stall indefinitely during its constraint solving.

Creating the specialized grammar and constraints for the subset of the ONNX MLIR dialect with only 25 operations required an in-depth understanding of the MLIR dialect and ISLa's specification language. This entailed at least 5 hours for a graduate student (the first author) to write the grammars and constraints. Further several days of effort were required for the graduate student to debug the constraints by learning the nuances of the ISLa specification language.
Further extending this by-hand to all four MLIR projects with 7 dialects and a total of 1,493 unique operations would require a tremendous manual effort. This motivates \tool{}'s approach to automate synthesis of custom mutations from existing tests to lower input constraint specification effort.

\subsection{Generalizability beyond MLIR}
We assess \tool{}'s ability to generalize beyond MLIR dialects by using \tool{} to generate AWS CloudFormation (CFN) templates. AWS CloudFormation templates are embedded in the JSON format, but keys and values have their own semantics. Since a refined grammar for CFN templates does not exist, we used an existing JSON grammar \cite{githubAntlrGrams} instead. The problem of generating CFN templates using the JSON grammar and existing tests is analogous to the problem of generating MLIR dialect specific inputs using the base MLIR grammar and existing tests. For our seed corpus, we extract 557 CFN templates from the PIPr dataset \cite{pipr, iacanalysis} that are accepted as valid by CloudFormation's linter, \texttt{cfn-lint} \cite{githubCfnLint}.

We compare \tool{} against Grammarinator to generate 2,000 CFN templates with the JSON grammar and the seed corpus of CFN templates. For each fuzzer, we measure the number of generated CFN templates that are accepted as valid by \texttt{cfn-lint}.

We found that 62.1\% (1,241 out of 2,000) CFN templates generated by \tool{} were valid, in comparison to 25.2\% (504 out of 2,000) CFN templates generated by Grammarinator. This $2.46\times$ improvement shows that \tool{}'s custom mutation synthesis generalizes beyond MLIR dialects to AWS CloudFormation.

\subsection{Threats to validity}

\subsubsection{Limited fuzzing time} In our experiments on code and dialect coverage, we limit the fuzzing budget to four hours for each fuzzer. While unlikely, continuing the fuzzing campaign for longer may reveal different trends.
\subsubsection{Choice of Subject Programs} To minimize bias, we selected four MLIR projects to represent a wide variety domains among 40 possible public MLIR projects. P1 (the LLVM/MLIR project) was chosen as it contains the original  MLIR core dialects that MLIRSmith defines its custom generators for. P2 is a deep-learning compiler for ONNX models that NeuRI directly fuzzed. P3 is a compiler for the Triton language. P4 (CIRCT) is a novel application of MLIR to the domain of hardware accelerator synthesis. 

\section{Related Work}
\label{sec:related}

\noindent \textbf{Grammar-based fuzzing.} Grammar-based fuzzers, e.g. Grammarinator~\cite{grammarinator}, Nautilus~\cite{aschermann2019nautilus}, LangFuzz~\cite{holler2012fuzzing} constrain input space with a context-free grammar. PolyGlot~~\cite{polyglot} transforms the high-level languages into a general IR with the BNF grammar given by the users and uses constraint mutators to preserve the grammar. 
Our experiments found that 97\% of Grammarinator's generated inputs fail to satisfy semantic constraints, e.g., type consistency, or more complex relationships over shapes and types.
\tool{}, instead, automatically infers semantic constraints from existing tests through mutation synthesis.

\noindent \textbf{Custom generator and mutation-based fuzzers.}
Generator-based fuzzers~\cite{padhye2019semantic,padhye2019jqf} require handwritten generators, usually coded in an imperative language, to produce valid inputs. 
CSmith~\cite{csmith} is a well-known example custom generator that produces random C programs for fuzzing C compilers. Related to our target domain of MLIR, MLIRSmith~\cite{mlirsmith} is a hand-written generator for  MLIR core dialects. Hand-coded generators encode context-sensitive constraints beyond the expressiveness of a context-free grammar. Despite the additional human effort required, MLIRSmith underperforms \tool{}, as it is unable to satisfy the input constraints of each new dialect beyond the MLIR core dialects. To target a new dialect, \tool{} does not require hand-coding custom mutators, as it infers parameterized, context-dependent mutations from example tests. 

Several studies have proposed fuzzers targeted at specific domains. Apart from fuzzing MLIR dialects, GrayC~\cite{grayc} designed custom mutators for fuzzing C programs. NNSmith~\cite{nnsmith} designed a generator for computation graphs to fuzz deep learning compilers.
BigFuzz~\cite{zhang2020bigfuzz} was proposed for Apache Spark programs, Qdiff~\cite{wang2021qdiff} for quantum programs,  HeteroFuzz~\cite{zhang2021heterofuzz} for heterogeneous applications with FPGA high-level-synthesis.
To effectively generate inputs, these fuzzers employ custom mutators. These mutators are hand-crafted and manually implemented for each domain. The amount of work needed to develop a new fuzzer for each domain highlights the need for \tool{} that synthesizes custom mutators from existing tests. These approaches do not allow for parameterized mutations whose content is adapt to the target context. The closest work to \tool{} is NeuRI~\cite{neuri}, which infers constraints over tensor shapes to generate deep learning (DL) models for fuzzing DL compilers. However, because it is specific to DL compilers and parameterization is limited to tensor shapes and operation's numerical attributes, NeuRI does not achieve high coverage in our experiments for MLIR.

\noindent \textbf{Learning constraints.}
ISLA~\cite{steinhofel2022input} allows semantic constraints to be applied during grammar-based fuzzing. Dewey et al.~\cite{dewey2014language} uses Constraint Logic Programming to specify constraints for test generation. 
As discussed in Section~\ref{sec:isla}, these techniques require significant human effort for specifying input constraints by hand~\cite{dewey2014language} or templates~\cite{steinhofel2022input}.

\noindent \textbf{Learning code patterns and transformations.} 
Code pattern inference techniques are used for code search~\cite{sivaraman2019active}, mining rules for detecting bugs~\cite{kang2021active}, and code quality~\cite{garg2022synthesizing}.
Other techniques synthesize patches, or program  transformation.
These patch synthesis techniques~\cite{meng2013lase, andersen2012semantic, serrano2020spinfer, lamothe2020a3, rolim2017learning, haryono2022androevolve, jiang2019inferring} transform programs by learning reusable transformation from examples. 
They identify parameterized patches given examples of the transformation, based on the observation that code elements re-occurring across multiple patches are essential.
\tool{} draws inspiration from these patch synthesis techniques by learning parameterized mutations from existing tests. 
While these techniques aim to mutate a buggy program into a single, correct program, \tool{} aims to diversify mutations for fuzzing.

\section{Conclusion}
\label{sec:conclusion}
Domain-specific fuzzers or test generators require months of development effort, which is impractical given the rapid evolution of target languages and intermediate representations. 
We present \tool{}, a novel approach that combines mutation synthesis with grammar-based fuzzing to make it easier to instantiate a domain-specific fuzzer without hand-coded customization and without manually specifying input constraints.

\tool{} synthesizes parameterized, context-dependent mutations from existing test cases, exploiting the observation that domain-specific input constraints are implicitly encoded in existing tests.
In our evaluation on MLIR-based compilers, \tool{} outperforms existing grammar-based and domain-specific fuzzers in terms of branch coverage by \avgCovImprove{} and dialect coverage by \avgDivImprove{}. \tool{} is able to synthesize custom mutations for new MLIR dialects. It also has the potential to generalize to different domains. For example, \tool{} improved the input validity rate from 25\% to 62\% when generating AWS CloudFormation templates.

A replication package has been made available at \url{https://github.com/UCLA-SEAL/SynthFuzz}.

\section*{Acknowledgment}
This work is supported by the National Science Foundation under grant numbers 2106838, 1764077, 1956322, and 2106404. It is also supported in part by funding from Amazon and Samsung.
We would like to thank the anonymous reviewers for their constructive feedback to help improve this work.

\bibliographystyle{IEEEtran}
\bibliography{main}

\end{document}